# A WEARABLE BREATH SENSOR BASED ON NATURAL CLINOPTILOLITE

G. Carotenuto, Institute for Polymers, Composites and Biomaterials - National Research Council. Piazzale E. Fermi, 1 – 80055 Portici (NA). Italy.

Low-power a.c. generators of square-wave or sinusoidal signals can be used in combination with impedimetric sensors to detect stimuli on the basis of the voltage drop taking place at the sensor electrodes. When a.c. generators with a power of only a few μ-Watts are used, this approach becomes extremely sensitive. A very low-power generator is the LCD back panel driving signal, which has a flipping polarity with a voltage of 3-5$V_{pp}$, depending on the generator model. This type of square-wave generator is contained in many low-cost handheld digital multimeters, and it is used as signal tracer to test, for example, low-frequency amplifiers. As an example, this method has been used to acquire a human breath rate pattern, by using a zeolite-based water sensor. If the generator I-V characteristics has been measured, the achieved breath pattern can be converted from a voltage drop vs. time graph to an impedance or current intensity vs. time graph.

Wearable technology [1] requires a special type of electronics to operate sensors [2]. Small size, lightness, and low-cost are the main characteristics of circuitry managing signals coming from wearable sensors [3]. In general, the adopted solutions are based on a very simplified electronic design, including only essential components.

Voltage is the simplest electrical property to be measured, consequently the possibility to use a root-mean-square (RMS) digital voltmeter [4,5] to determine the wearable sensor response represents a very desirable solution. Usually, the response to stimuli of an impedimetric sensor is detected as impedance (Z) variation, measured by a LCR-meter, or as variation of the intensity of current flowing in the device [6,7]. However, if the impedimetric sensor is biased by a very low power generator, the sensor response can be detected also as voltage drop at sensor electrodes. In particular, lower is the generator power, higher is the sensitivity of this method. In addition, the I-V characteristics of the generator can be determined and the voltage response can be converted to current intensity or impedance variation by using this generator I-V characteristics.

Recently, a breath rate sensor based on natural clinoptilolite has been proposed [8], the breath pattern was generated by measuring the current intensity flowing at sample surface after biasing this sensor with a 5kHz sinusoidal voltage signal (20$V_{pp}$). However, a similar pattern with much higher resolution can be generated by using the here proposed electrical method. To the best of our knowledge, such instrumental approach, based on the combination of a low-power a.c. generator with a true-RMS digital voltmeter, has been proposed here for the first time to detect the response of an impedimetric ceramic sensor like, for example, a zeolite water sensor.

When a zeolitic sensor is biased by a sinusoidal or square-wave voltage signal, produced by an ordinary a.c. source (function generator with power >1W), the very high impedance (few MΩ) of the zeolitic sensor can cause just a negligible voltage drop, when the stimulus (breathing) is detected by the sensor. However, many digital multimeters (DMMs) incorporate a square-wave source (signal tracer) of very low power (of the μW order), and if these generators are used to bias the sensor, a significant voltage drop can be experienced even for a slight variation of the high load impedance. Such type of signal generators can be used to detect stimuli by an impedimetric zeolitic sensor by measuring the voltage drop at the sensor electrodes. In this work the square-wave source of several handheld DMMs has been tested (DT830D, DT-830B, DT832, ANENG AN8206, ANENG AN8008, KONIG KDM-100, UT20B) and the best results (lowest power) have been obtained by the DT830D multimeter model. An alternated voltage source is required to avoid ion accumulation at electrodes surface. In particular, the square-wave output incorporated in the DT830D multimeter is based on the LCD driver in the multimeter 7106 chip. The integrated 7106 is an ADC, which provides also the LCD back panel driving signal with flipping polarity.

Two electrical contacts were printed on the surface of a zeolite slab by using silver paint (ENSON, EN-06B8). The zeolite sensor was connected to the square-wave output and a true-RMS digital voltmeter (UNI-Trend, UT71D multimeter) was placed in parallel to the generator, according to the electrical scheme shown in Figure 1. The square-wave signal characteristics are show in Figure 2(a,b). As visible in the oscillogram, when a coupling capacitor of 10nF was used to filter the small offset voltage contained in the signal, the power source gave a perfectly symmetrical square-wave with an amplitude of 5.0$V_{pp}$ (for a square-wave signal, the effective voltage, $V_{eff}$, measured by a true-RMS digital voltmeter is exactly corresponding to the peak voltage value, $V_p$=2.5). The harmonic composition of the square-wave signal was obtained by a FFT analysis based on a 2 channel 10MHz USB oscilloscope (PicoScope 2204-D2). As visible in Figure 2b, the square-wave signal analysis in the frequency domain showed that the signal was composed by the fundamental at the frequency of 50Hz and three main odd harmonics.

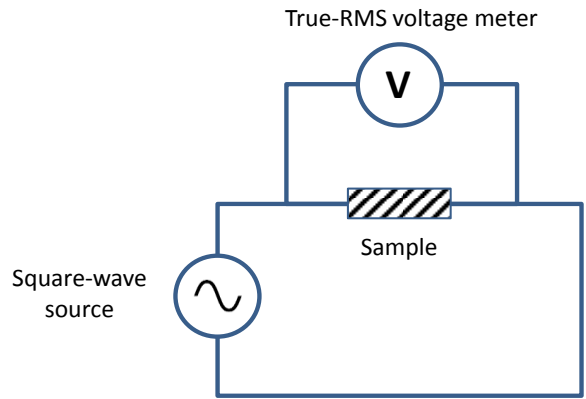

**Fig. 1** – Electrical circuit used for the voltage drop measurements.

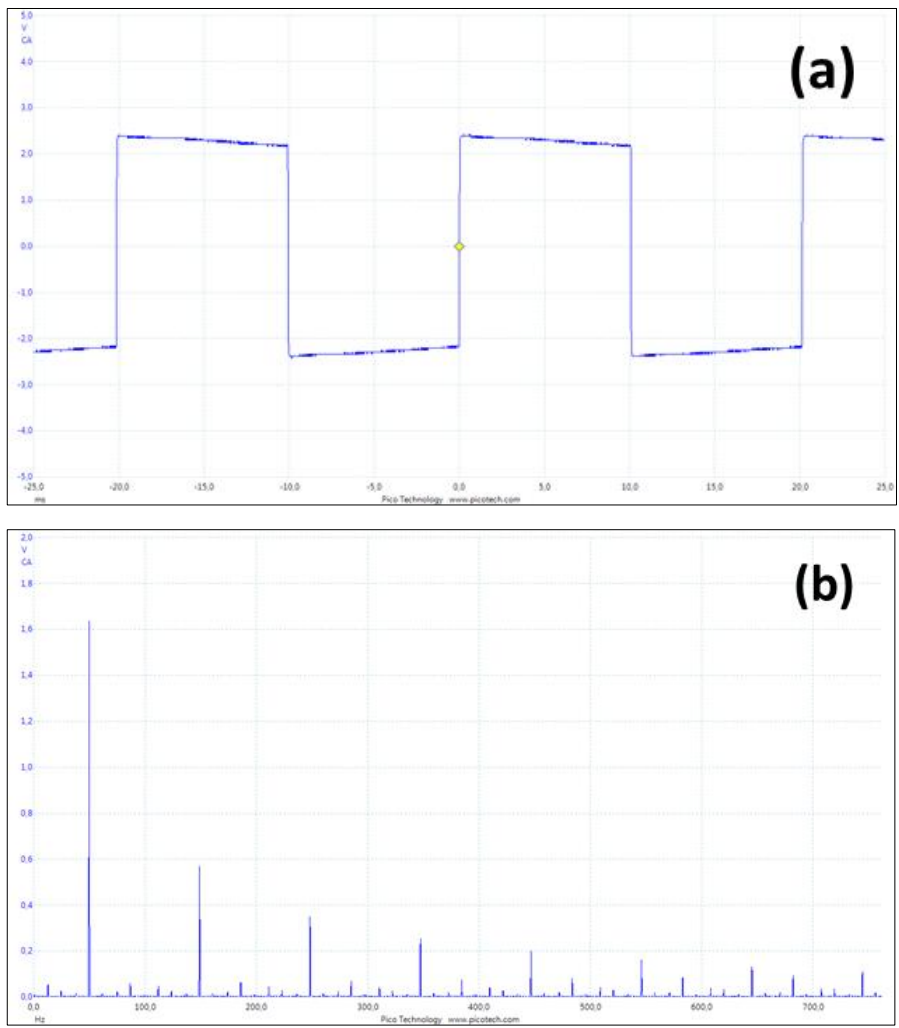

**Fig. 2** – Square-wave signal (a) and its harmonic analysis by FFT oscilloscope (b).

The true-RMS voltage was recorder vs. time at 8-9Sa/s during the exposition to human breathing by using the high speed data logger incorporated in the digital voltmeter UT71D, set at low resolution (when the digital multimeter is set at 4,000 count mode, fast logging is allowed). An example of breathing pattern, including three breathing events, obtained by the voltage drop technique, is shown in Figure 3. According to this breathing pattern, the surface resistivity was quickly modified in presence of the water vapor, and the voltage decreased of ca. 100mV. As visible, water adsorption was a very fast process and followed a linear temporal behavior, while desorption was slightly slower and followed a parabolic behavior.

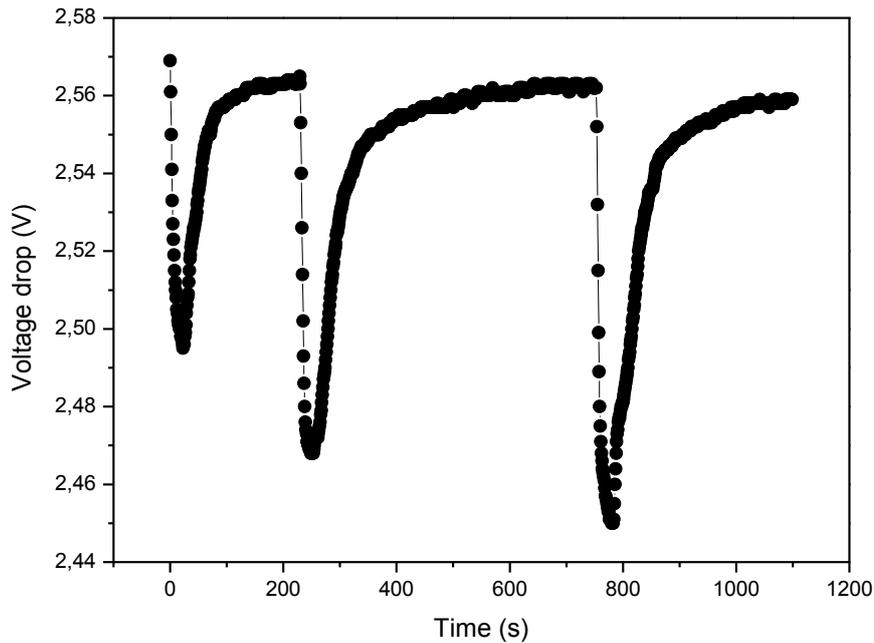

**Fig. 3** – Human breathing pattern obtained by the voltage drop technique.

The I-V characteristics of the square-wave source is shown in Figure 4a. This curve has been obtained by applying to the generator a gradually increasing pure resistive load of a precisely known value (tolerance of resistors: ±1%), and measuring the output voltage and the absorbed current by true-RMS digital voltmeter and digital ammeter, respectively. A resistance decade box (PeakTech 3280) was used for this purpose. The generator I-V characteristics clearly shows a non-ideal behavior with an electromotive force (e.m.f.) of ca. 2.57V, and an internal resistance value (ratio between the e.m.f. and $I_{max}$) of ca. 120kΩ. In particular, the short-circuit voltage (V=0 and I=$I_{max}$) and the open-circuit voltage, that is, the voltage without load (V=$V_{max}$ and I=0), have been obtained by extrapolation, since direct measurement of the short-circuit voltage could damage the generator for the high current intensity flowing in the circuit (overload protection is not present in such simple type of generators), while the open-circuit voltage is limited by the digital multimeter input impedance (usually 10MΩ at 50Hz). In order to avoid the influence of the digital voltmeter input impedance, voltage and current values were measured for resistive loads inferior to 10MΩ.

As visible in Figure 4b, the power curve of the square-wave generator shows a dependence of the output power on the applied resistive load, with a maximum value of ca. 14µW. Such maximum power is achieved with a pure resistive load of 120kΩ, that corresponds exactly to the internal resistance of the generator.

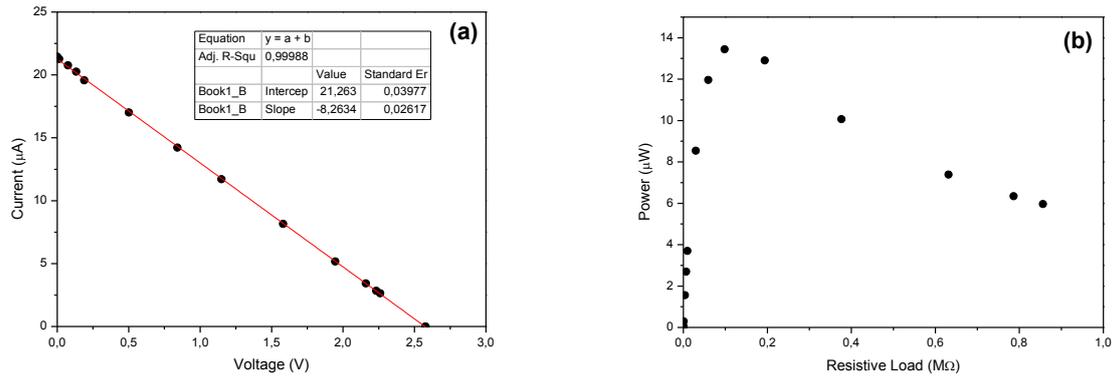

**Fig. 4** – I-V characteristics of the square-wave generator (a) and its corresponding power curve (b).

As indicated above, the breathing pattern displayed as voltage drop (see Figure 3) can be converted to a true-RMS current intensity variation or to a normalized impedance variation by using the measured I-V characteristics of the square-wave generator (see Figure 5). In particular, the following relationship has been used to convert the voltage drop vs. time curve to a current vs. time curve:

$$I = I_{cc} \cdot (1-V/E) \qquad (1)$$

where $I_{cc}=I_{max}=21.3\mu A$ is the short circuit current intensity and E is the e.m.f.. Similarly, the zeolite sensor impedance (Z) vs. time curve can be obtained from the voltage vs. time data by the following relationship:

$$Z = V/I = V/[I_{cc} \cdot (1-V/E)] = V/[I_{cc}-(1/r) \cdot V] \qquad (2)$$

where r=120kΩ is the generator internal resistance. As visible, when the breath pattern is expressed as a temporal variation of the impedance a slight distortion of the peaks can be observed (see Figure 5b).

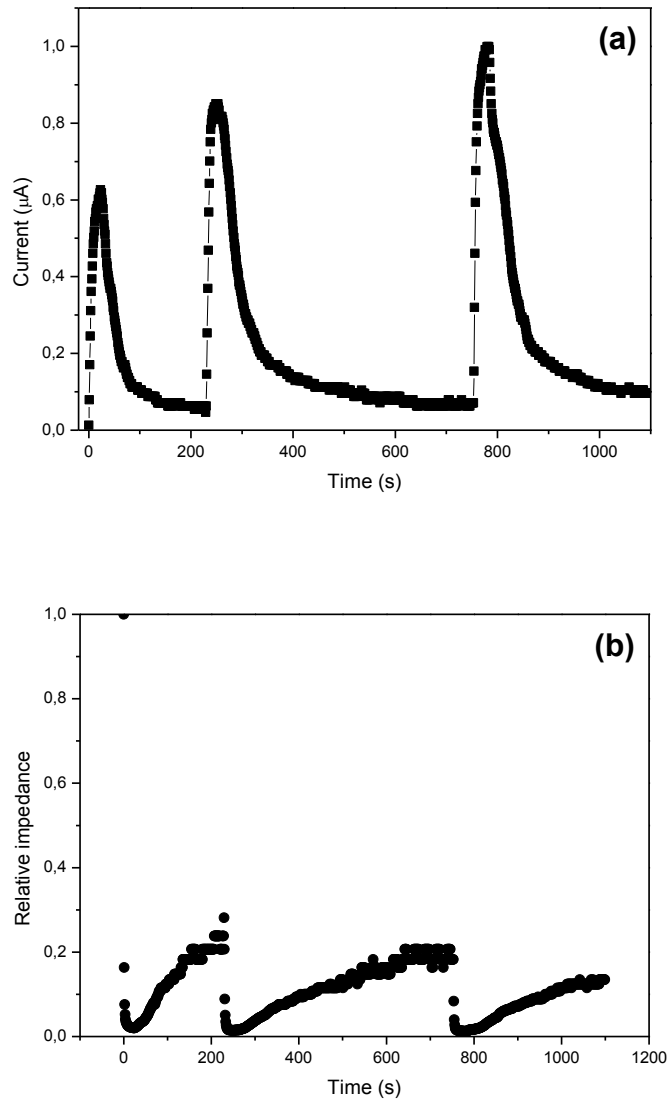

**Fig. 5** – Signals displayed as current (a) and relative impedance (b) vs. time.

In conclusion, the above described technique to detect conductivity changes in an impedenziometric sensor based on the measurement of the voltage drop represents a simple, very sensitive, and economically convenient approach to monitor the signals. This method can be applied to a wearable sensor like, for example, a zeolite-based wearable breath sensor, since it is based on the use of a small, inexpensive, and lightweight signal source and a true-RMS digital voltmeter. The main limitation of this technique has been the use of the quite low frequency (50Hz or 60Hz) typically used for signal tracers, which determined a high value of the material impedance for the presence of a reactive part. There are digital multimeters with an output frequency that can be varied by steps (e.g., ANENG AN8008), however such digital multimeters have power values higher than the simple DT830D and therefore they involve lower voltage drop during sensing.